\documentclass[12pt]{iopart}
\usepackage{amssymb}
\usepackage{graphicx}
\usepackage{color}

\def\beq{\begin{eqnarray}}
\def\eeq{\end{eqnarray}}

\begin{document}

\title{High--order connected moments expansion for the Rabi Hamiltonian}

\author{Paolo Amore\footnote{Corresponding author}}
\address{Facultad de Ciencias, CUICBAS, Universidad de Colima, \\
Bernal D\'{\i}az del Castillo 340, Colima, Colima, Mexico}
\author{Francisco M. Fern\'andez}
\address{INIFTA (UNLP,
CCT La Plata-CONICET), Divisi\'{o}n Qu\'{i}mica Te\'{o}rica, Diag.
113 y 64 (S/N), Sucursal 4, Casilla de Correo 16, 1900 La Plata,
Argentina}
\author{Martin Rodriguez}
\address{Facultad de Ciencias, Universidad de Colima, \\
Bernal D\'{\i}az del Castillo 340, Colima, Colima, Mexico}


\begin{abstract}
We analyze the convergence properties of the connected moments
expansion (CMX) for the Rabi Hamiltonian. To this end we calculate
the moments and connected moments of the Hamiltonian operator to a
sufficiently large order. Our large--order results suggest that
the CMX is not reliable for most practical purposes because the
expansion exhibits considerable oscillations.
\end{abstract}


\maketitle


\section{Introduction}

\label{sec:intro}

Some time ago Horn and Weinstein\cite{HW84} proposed a systematic
nonperturbative technique for the calculation of ground--state expectation
values of arbitrary operators. It is based on the well--known expansions in
terms of cumulants or semi--invariants\cite{C46,K62} of quantum--mechanical
expectation values in which the exponential operator $e^{-t\hat{H}}$ takes
place. Although the theoretical results are rigorous there remains the
practical problem of summing the resulting $t$--expansion in order to obtain
the desired expectation value in the limit $t\rightarrow \infty $. They
resorted to Pad\'{e} approximants and later Stubbins\cite{S88} proposed
other ways of extrapolating the $t$--series. However, those results were not
encouraging.

Cioslowski\cite{C87a} proposed a clever extrapolating technique based
on a series of exponential functions and derived an appealing expression
that has become popular as the connected--moments expansion (CMX). Later
Knowles\cite{K87} derived a more systematic way of obtaining the CMX.

The CMX results on the H$_2$ molecule appeared quite promising; however it
seems that the promised test of the CMX on multideterminant wave functions%
\cite{C87a} has never been published. Knowles\cite{K87} showed that although
the initial terms of the CMX recover a large fraction of the correlation
energy in molecular calculations, subsequent terms converge to an incorrect
energy.

The CMX is quite appealing because it provides approximate values for the
ground--state energy of a quantum system directly in terms of a finite
number of connected moments. This may be the reason why the CMX and its
variants\cite{K87,S88,MZM94} were applied to several simple physical problems%
\cite{C87a,K87,MBM89,MPM91,C87b,C87c,C87d,C87e,FMB02,FCMMM10} in
spite of its
limitations\cite{K87,S88,MZM94,MBM89,MPM91,MZMMP94,LL93, WS99}.

In order to overcome some of the drawbacks of the CMX several
authors have proposed alternative strategies like the generalized
moment expansion (GMX) \cite{MMFB05,FMMB06,FMBC08}.
Bartashevich\cite{B08} proposed the connected--moments polynomial
approach that yields approximate eigenvalues for all states as
roots of a simple polynomial function of the energy with
coefficients that depend on the moments of the Hamiltonian
operator. This approach was later proved to be equivalent to the
Rayleigh--Ritz variational method\cite{MD33} in the Krylov
space\cite{M08,F08,F09} that we will call RRK from now on.
Numerical experiments proved that the RRK converges more smoothly
and is therefore more reliable than the CMX when both methods are
applied to the simple models so far chosen for testing the
latter\cite{F09,AF09b}.

Some time ago Fessatidis et al\cite{FMB02} applied the CMX and one
of its variants, the alternative moments expansion (AMX), to a
non--trivial problem with many physical applications: the Rabi
hamiltonian. Because they only considered low order expansions
there is no clear indication about the convergence of the moments
expansions for that important model. The purpose of this paper is
to investigate the convergence of the CMX for the Rabi Hamiltonian
numerically. Such analysis requires moments expansions of
sufficiently large order for different values of the model
parameters. We expect that the conclusions drawn for the Rabi
Hamiltonian may be of utility for future applications of the CMX
and its variants to more realistic physical problems.

The paper is organized as follows: in Section \ref{sec:Rabi} we introduce
the Rabi Hamiltonian and discuss the diagonalization of its matrix in an
appropriate basis set; in Section \ref{sec:Cumulant} we outline the $t$%
--expansion; in Section~\ref{sec:CMX} we outline the main equations of the
CMX and compare its results with those obtained by means of the RRK and the
accurate diagonalization. Finally in Section \ref{sec:comments} we draw
conclusions.


\section{The Rabi Hamiltonian: exact diagonalization}

\label{sec:Rabi}

The Rabi Hamiltonian is a model of a two level atom or spin system coupled
to a single--mode bosonic field given by the Hamiltonian operator
\[
\hat{H}=\frac{1}{2}\omega _{0}\sigma _{z}+\omega \hat{b}^{\dagger }\hat{b}%
+g\left( \sigma _{+}+\sigma _{-}\right) \left( \hat{b}^{\dagger }+\hat{b}%
\right) \ .
\]
where $\sigma _{i}$ are the well known Pauli matrices $\omega $
and $\omega _{0}$ are the physical parameters that determine the
spectrum in absence of coupling, and $g$ is the coupling between
the atom and the bosonic field. When $g=0$ the spin and bosonic
degrees of freedom decouple and the problem is exactly solvable.
For this reason it is expected that any approach yields better
results for small values of $g$. Recently Pan and
coworkers~\cite{PGWD10} have shown that the Rabi hamiltonian can
be solved almost exactly using a progressive diagonalization
scheme.

Although this model is not exactly solvable for $g\neq 0$, one can easily
obtain highly accurate numerical results by, for example, straightforward
diagonalization of the Hamiltonian matrix in an appropriate basis set. Since
we will need such results in our analysis of the performance of the CMX, we
proceed to describing the diagonalization procedure. The Hilbert space for
this problem is spanned by a basis set of states given by the direct product
of the spin and bosonic ones. We label them in the following way:
\[
|n\rangle =\left\{
\begin{array}{ccc}
|\downarrow \rangle \otimes |\left[ \frac{n-1}{2}\right] \rangle & , & n\ odd
\\
|\uparrow \rangle \otimes |\left[ \frac{n-1}{2}\right] \rangle & , & n\ even
\end{array}
\right. \ ,
\]
where $n=1,2,\dots $ and $\left[ a\right] $ means integer part of $a$.

\begin{figure}[tbp]
\bigskip
\par
\begin{center}
\includegraphics[width=6cm]{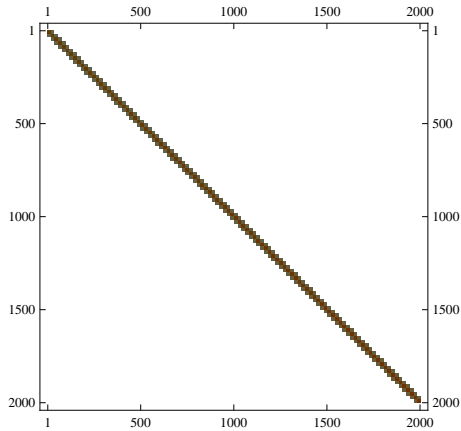}
\end{center}
\caption{Matrix plot of the matrix corresponding to the Rabi Hamiltonian for
$\omega_0=\omega=1$ and $g=5$, with $N=2000$.}
\label{Fig_1}
\end{figure}

The calculation of the matrix elements of the operators that are relevant
for the model is straightforward; for example:
\begin{eqnarray}
B_{nm} &\equiv &\langle n|\hat{b}|m\rangle =\left\{
\begin{array}{ccc}
\sqrt{\left[ \frac{m-1}{2}\right] } & , & \left[ \frac{n-1}{2}\right]
=\left[ \frac{m-1}{2}\right] -1 \\
0 & , & otherwise
\end{array}
\right. \ ,  \nonumber \\
\left( \Sigma _{z}\right) _{nm} &\equiv &\langle n|\hat{\sigma}_{z}|m\rangle
=\left\{
\begin{array}{ccc}
(-1)^{n} & , & n=m \\
0 & , & otherwise
\end{array}
\right. \ ,  \nonumber \\
\left( \Sigma _{+}\right) _{nm} &\equiv &\langle n|\hat{\sigma}_{+}|m\rangle
=\left\{
\begin{array}{ccc}
2 & , & m\ odd\ and\ n=m+1 \\
0 & , & otherwise
\end{array}
\right. \ ,  \label{eq:mat_el}
\end{eqnarray}
and those for $\hat{b}^{\dagger }$ and $\sigma _{-}$ are the
hermitian conjugates of the matrices for $\hat{b}$ and $\sigma
_{+}$. The matrix for the Rabi Hamiltonian follows from obvious
straightforward products of those matrices. If we restrict the
space to the first $N$ states then we obtain an $N\times N$ matrix
$\mathbf{\ H}$ that we can diagonalize numerically in order to
obtain approximate eigenvalues and eigenvectors. The calculation
is
greatly facilitated by the fact that the matrix $\mathbf{H}$ is \textsl{%
sparse}; i.e. most its matrix elements are zero. This property is quite
useful from a computational point of view because it allows one to save
computer memory. Fig.\ref{Fig_1} shows a simple graphical representation of
the matrix $\mathbf{H}$ for $N=2000$, $\omega _{0}=\omega =1$ and $g=5$. It
is worth noting that the truncation of the Hilbert space preserves the
symmetry of the problem determined by the commutation of the operators $\hat{%
H}$ and $\hat{\Pi}=e^{i\pi \hat{n}}$, where $\hat{n}=\hat{b}^{\dagger }\hat{b%
}+\frac{1}{2}+\frac{1}{2}\sigma _{z}$ (the corresponding matrices also
commute).

Table \ref{tab1} shows the ground--state energy of the Rabi
Hamiltonian for several values of $N$ and for the same set of
parameters chosen by Bishop et al\cite{BDQV99}. Present results
are more accurate than those obtained earlier and will be a useful
benchmark for the investigation of the convergence properties of
the CMX. We have calculated the matrix $\mathbf{H}$ analytically
in terms arbitrary model parameters so that we do not have to
calculate it again each time that we decide to modify those
parameters. This strategy makes the calculation quite efficient.

We have calculated the lowest eigenvalues of the matrices by
iterative application of the conjugate gradient method (CGM) to
the numerical matrix $\mathbf{H}$. Notice that our results are
fully converged for a $1000 \times 1000$ matrix, except in the
last case considered where just the last digit is not correct.

\begin{table}[tbp]
\caption{Ground state energy of the Rabi Hamiltonian for specific values of
the parameters. $N$ is the number of states used in the exact
diagonalization. The results may be compared with the variational results of
ref.\protect\cite{BDQV99}.}
\label{tab1}%
\begin{tabular}{ccccc}
$N$ & $\omega_0$ & $\omega$ & $g$  & $E_{1}$  \\
\hline
1000 & 1 & 1 & 5 & -100.000626570374178204295743532860688625653449650 \\
1500 & 1 & 1 & 5 & -100.000626570374178204295743532860688625653449650 \\
2000 & 1 & 1 & 5 & -100.000626570374178204295743532860688625653449650 \\
ref.\cite{BDQV99} & 1 & 1 & 5 & -100.001 \\
\hline
1000 & 1 & 2 & 5 & -50.001262757900797977214814102896046471301336432 \\
1500 & 1 & 2 & 5 & -50.001262757900797977214814102896046471301336432 \\
2000 & 1 & 2 & 5 & -50.001262757900797977214814102896046471301336432 \\
ref.\cite{BDQV99} & 1 & 2 & 5 & -50.0013 \\
\hline
1000 & 2 & 1 & 5 & -100.002506281526606167493915731790865439561963879 \\
1500 & 2 & 1 & 5 & -100.002506281526606167493915731790865439561963878 \\
2000 & 2 & 1 & 5 & -100.002506281526606167493915731790865439561963878 \\
ref.\cite{BDQV99} & 2 & 1 & 5 & -100.003 \\
\end{tabular}
\bigskip\bigskip
\end{table}


\section{The cumulant or $t$--expansion}

\label{sec:Cumulant}

For concreteness, in this section we outline the main ideas behind the
so--called $t$--expansion (or cumulant expansion)\cite{HW84}. The
moment--generating function
\begin{equation}
Z(t)=\left\langle \varphi \right| e^{-t\hat{H}}\left| \varphi \right\rangle
=\sum_{j=0}^{\infty }\frac{(-t)^{j}}{j!}\mu _{j}
\end{equation}
gives us the moments $\mu _{j}=\left\langle \varphi \right| \hat{H}%
^{j}\left| \varphi \right\rangle $ of the hamiltonian operator $\hat{H}$,
where $\left| \varphi \right\rangle $ is a properly chosen trial state. The
expectation value of $\hat{H}$ in the state $e^{-t\hat{H}/2}\left| \varphi
\right\rangle $
\begin{equation}
E(t)=-\frac{Z^{\prime }(t)}{Z(t)}=\frac{\left\langle \varphi \right| \hat{H}%
e^{-t\hat{H}}\left| \varphi \right\rangle }{\left\langle \varphi \right|
e^{-t\hat{H}}\left| \varphi \right\rangle }
\end{equation}
exhibits several interesting properties: first, $E(t)\geq E_{0}$, where $%
E_{0}$ is the ground--state energy, second, $E^{\prime }(t)\leq 0$ and,
third, $\lim_{t\rightarrow \infty }E(t)=E_{0}$ provided that the overlap
between $\left| \varphi \right\rangle $ and the ground state $\left| \psi
_{0}\right\rangle $ is nonzero.

The function $E(t)$ is closely related to the cumulant function
$K(t)$ defined by $Z(t)=e^{K(t)}$.\cite{C46,K62} The formal Taylor
series of $E(t)$ about $t=0$ yields the $t$--expansion:
\begin{equation}
E(t)=\sum_{j=0}^{\infty }\frac{(-t)^{j}}{j!}I_{j+1}  \label{eq:t-exp}
\end{equation}
where the cumulant coefficients (or connected moments) $I_{j}$ can be easily
obtained from the recurrence relation
\begin{equation}
I_{j+1}=\mu _{j+1}-\sum_{i=0}^{j-1}\left(
\begin{array}{c}
j \\
i
\end{array}
\right) I_{i+1}\mu _{j-i}  \label{eq:I_j}
\end{equation}
The main advantage of the methods based on the cumulant or
connected--moments expansion is that they are size
extensive\cite{HW84,K87}.

Any practical application of this method requires a suitable
extrapolation of the $t$--expansion (\ref{eq:t-exp}) to
$t\rightarrow \infty $ in order to obtain $E_{0}$. This is not a
simple task and different extrapolation techniques may lead to
different results. As we have already mentioned before the first
application of the $t$--expansion was based on Pad\'e approximants
that provide the simplest and most straightforward strategy\cite
{HW84}. At this point we want to point out a common misconception
about the theorem of Horn and Weinstein\cite{HW84}. Many authors
state that the function (\ref{eq:t-exp}) converges for
$t\rightarrow \infty $\cite{C87a,MBM89,MPM91,SKMT97,LL93,FMMB06}
which is not the case as one can easily verify. For some complex
values of $t$ the function $Z(t)$ may vanish and therefore the
$t$--expansion converges for $t<|t_s|$ where $t_s$ is the singular
point of $E(t)$ closest to the origin in the complex $t$--plane.
This fact has already been discussed by Witte and
Shankar\cite{WS99} and one may easily convince oneself that it is
so by means of the exactly solvable two--dimensional model
discussed by Knowles\cite{K87}.

Note that the expansion in terms of exponential functions of $t$
\begin{equation}
E(t)=E_{0}+\sum_{j=1}^{\infty }A_{j}\exp (-b_{j}t),
\label{eq:exp-exp}
\end{equation}
which is the basis of the CMX\cite{C87a}, does not take into
account the singular points of $E(t)$ and therefore the matching
of the $t$--expansion is only valid for $t<|t_{s}|$. Consequently,
it is unlikely that we can successfully
extrapolate the expression (\ref{eq:exp-exp}) thus derived to the limit $%
t\rightarrow \infty $.

Regardless of which extrapolation scheme we may use, it is clear that the
application of the theorem proved by Horn and Weinstein requires the
calculation of a certain number of connected moments of the Hamiltonian. For
problems of great complexity, such as for example many--body or quantum
field theory problems, it may be quite difficult (or even impossible) to
carry out this task for sufficiently large (or even modest) orders. In such
cases one should therefore rely on the extrapolation of the expansion with a
few connected moments. It is not easy to prove the accuracy of such
extrapolations in the general case, and for this reason it is useful to
verify what happens in the case of a simple though non--trivial model, like
the Rabi Hamiltonian, where it is possible to obtain large--order moments
and sufficiently accurate numerical results. We expect that a careful
investigation of the convergence properties of the connected moments
expansions for this model will then serve as a guide for others in which one
cannot carry out calculations to such large orders.

We first outline the procedures that we follows for the
calculation of exact analytical moments of the Hamiltonian. In the
first place, we resort to an $N\times N$ matrix representation
$\mathbf{H}$ of the Hamiltonian operator,
as a function of the parameters of the model. for simplicity we assume that $%
N$ is even, i.e. that we are working in a subspace of the Hilbert space
containing at most $N/2$ bosons. If $N$ is sufficiently large (for example, $%
N=2000$) we can safely calculate the first (say $100$) moments exactly. The
effect of the space truncation does not affect the calculation because of
the band form of the matrix shown in Fig.~(\ref{Fig_1})~\footnote{%
The Rabi Hamiltonian allows transitions which involve a change in the number
of bosons by one unit and a spin flip. This means that, as long as $N$ is
chosen large enough, the moments of the Hamiltonian in $|\Psi_0\rangle$ may
be obtained exactly.}. In this approximation the moment $\mu
_{j}=\left\langle \hat{H}^{j}\right\rangle $ is simply given by the
corresponding diagonal element of $\mathbf{H}^{j}$. We have also resorted to
the coordinate representation and calculated the moments analytically in
order to verify the accuracy of the matrix approach just outlined.

In order to calculate the moments we follow Fessatidis et
al~\cite{FMB02} and choose the trial state $|\varphi \rangle
=|\downarrow \rangle \otimes |0\rangle
=|1\rangle $ to be the ground--state of the noninteracting Hamiltonian ($g=0$%
). In this way we can readily calculate the moments of the Rabi Hamiltonian
systematically and analytically by straightforward matrix--matrix and
matrix--vector multiplications (or in the alternative way indicated above).
To this end we resorted to the symbolic operations provided by available
computer algebra software like Mathematica and wrote a code that produces
the desired moments and connected moments in a reasonably short time. On the
other hand, Fessatidis et al~~\cite{FMB02} only derived the first five
moments of the Rabi Hamiltonian, a fact which considerably limited the
accuracy of their results as well as the reliability of their conclusions.

Alternatively the moments of the Rabi hamiltonian may be calculated by representing
the Hamiltonian operator as
\begin{eqnarray}
\hat{H} = \left( \begin{array}{cc}
\frac{\omega_0-\omega}{2} - \frac{1}{2} \frac{d^2}{dx^2} + \frac{\omega^2x^2}{2} &  2\sqrt{2\omega} g x\\
2\sqrt{2\omega} gx & \-\frac{\omega_0+\omega}{2} - \frac{1}{2} \frac{d^2}{dx^2} + \frac{\omega^2x^2}{2} \\
\end{array} \right)
\end{eqnarray}
and writing the trial state as
\begin{eqnarray}
\Psi_0(x) = \left( \begin{array}{c}
0 \\
\left(\frac{\omega}{\pi}\right)^{1/4} e^{-\omega x^2/2} \\
\end{array} \right) \ .
\end{eqnarray}
In this way we can obtain moments of sufficiently large order by
straightforward differentiation and integration and compare them
with those provided by the matrix--vector procedure outlined
above.

With the purpose of comparison we show the first five moments
calculated in the two ways indicated above:
\begin{eqnarray}
\mu _{1} &=&-\frac{\omega _{0}}{2}  \nonumber \\
\mu _{2} &=&\frac{\omega _{0}^{2}}{4}+4g^{2}  \nonumber \\
\mu _{3} &=&g^{2}(4\omega -2{\omega _{0}})-\frac{{\omega _{0}}^{3}}{8}
\nonumber \\
\mu _{4} &=&48g^{4}+2g^{2}\left( 2\omega ^{2}+\omega _{0}^{2}\right) +\frac{%
\omega _{0}^{4}}{16}  \nonumber \\
\mu _{5} &=&8g^{4}(20\omega -3\omega _{0})+g^{2}\left( 4\omega ^{3}+2\omega
^{2}\omega _{0}+2\omega \omega _{0}^{2}-\omega _{0}^{3}\right) -\frac{\omega
_{0}^{5}}{32}\ .  \label{eq:mu_j}
\end{eqnarray}
It is worth noting that our moment $\mu _{3}$ is different from
the one shown in Eq.~(12) of the paper by Fessatidis et
al\cite{FMB02}. We have verified that this discrepancy was not
merely a misprint so that their results and conclusions may be
incorrect. We will come back to this point later on.

Once we have the moments the calculation of the connected moments
by means of Eq.~(\ref{eq:I_j}) is straightforward. The first five
ones are
\begin{eqnarray}
I_{1} &=&-\frac{\omega _{0}}{2}  \nonumber \\
I_{2} &=&4g^{2}  \nonumber \\
I_{3} &=&4g^{2}(\omega +\omega _{0})  \nonumber \\
I_{4} &=&4g^{2}(\omega +\omega _{0})^{2}  \nonumber \\
I_{5} &=&4g^{2}\left( (\omega +\omega _{0})^{3}-16g^{2}\omega _{0}\right) \ .
\label{eq:I_j_2}
\end{eqnarray}

\begin{figure}[tbp]
\bigskip
\par
\begin{center}
\includegraphics[width=10cm]{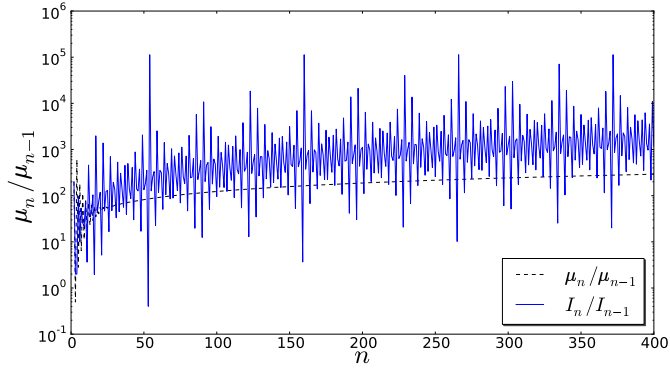}
\end{center}
\caption{Ratio of moments $\mu_{n}/\mu_{n-1}$
(dashed line) and of connected moments $I_{n}/I_{n-1}$ (solid line) for the Rabi Hamiltonian with $\omega
_{0}=\omega =1$ and $g=5$.}
\label{Fig_2a}
\end{figure}
Fig.~\ref{Fig_2a} shows the ratios $\mu _{n}/\mu _{n-1}$ and
$I_{n}/I_{n-1}$ for the Rabi Hamiltonian with $\omega _{0}=\omega
=1$ and $g=5$. Note that the ratio of the connected moments
exhibits an irregular oscillatory behaviour that suggests that we
may encounter difficulties in the summation of the $t$--series (in
fact, in this case the function $f(t)$ proposed by
Knowles\cite{K87} is not even close to a Stieltjes series).


\section{Connected--Moments Expansion}

\label{sec:CMX}

In order to apply the CMX to the Rabi Hamiltonian we resort to the
beautifully compact expression for the correlation energy $%
E_{corr}=E_{0}-I_{1}$ derived by Knowles:\cite{K87}
\begin{equation}
E_{corr}^{(m)}=\left(
\begin{array}{cccc}
I_{2} & I_{3} & \dots & I_{m+1}
\end{array}
\right) \ \left(
\begin{array}{cccc}
I_{3} & I_{4} & \dots & I_{m+2} \\
I_{4} & I_{5} & \dots & I_{m+3} \\
\dots & \dots & \dots & \dots \\
I_{m+2} & I_{m+3} & \dots & I_{2m+1}
\end{array}
\right) ^{-1}\ \left(
\begin{array}{c}
I_{2} \\
I_{3} \\
\dots \\
I_{m+1}
\end{array}
\right) \   \label{eq:E_corr_CMX}
\end{equation}
Knowles\cite{K87} also discussed the following alternative expression for
the correlation energy:
\begin{equation}
E_{corr}^{(m)}=-\frac{S_{21}^{2}}{S_{31}}\ \left( 1+\frac{S_{22}^{2}}{%
S_{21}^{2}S_{32}}\ \left( 1+\frac{S_{23}^{2}}{S_{22}^{2}S_{33}}\left(
1+\dots \left( 1+\frac{S_{2m}^{2}}{S_{2,m-1}^{2}S_{3m}}\right) \right)
\right) \right) \ ,  \label{eq:E_corr_2}
\end{equation}
where $S_{k1}=I_{k}$ ($k=2,3,\dots $) and $%
S_{k,i+1}=S_{ki}S_{k+2,i}-S_{k+1,i}^{2}$ developed by Cioslowski\cite{C87a}.
These two expressions are not equivalent and yield different series in
powers of the coupling constant $g$. In this paper we calculate the
correlation energy by means of Eq.~(\ref{eq:E_corr_CMX}) that appears to be
more accurate and suitable for present large--order calculations.

We have been able to obtain a sufficiently large number of moments and
connected moments of the Rabi Hamiltonian by means of the procedures
described above. The square matrix appearing in Eq.~(\ref{eq:E_corr_CMX})
may be badly conditioned and therefore its inverse may contain large
numerical errors unless its elements are known with sufficiently large
accuracy.

In this paper we have decided to compare the performance of the CMX with
that of the Rayleigh--Ritz variational method in the Krylov space (RRK). In
the latter approach we choose the nonorthogonal basis set $\left\{ \left|
\phi _{j}\right\rangle =\hat{H}^{j}\left| \varphi \right\rangle \right\}
_{j=0}^{\infty }$, where $\left| \varphi \right\rangle $ is a properly
chosen trial state. In this particular implementation of the Rayleigh--Ritz
variational method the secular equations become\cite{K87,F08,F09,AF09b}
\begin{equation}
\sum_{i=0}^{N-1}\left( \mu _{i+j+1}-W\mu _{i+j}\right)
c_{i}=0,\;j=0,1,\ldots ,N-1  \label{eq:secular}
\end{equation}
There are nontrivial solutions to the homogeneous system of linear equations
(\ref{eq:secular}) only for the $N$ values of $W=W_{0},W_{1},\ldots ,W_{N-1}$
that are roots of the secular determinant
\begin{equation}
\left| \mu _{i+j+1}-W\mu _{i+j}\right| _{i,j=0}^{N-1}=0.
\label{eq:secular_det}
\end{equation}
It is well known that the Rayleigh--Ritz approximate eigenvalues $W_{j}$
converge monotonously from above towards the actual eigenvalues $E_{j}$ of
the Hamiltonian operator $\hat{H}$. In particular, if $\left| \varphi
\right\rangle $ is not orthogonal to the ground state $\left| \psi
_{0}\right\rangle $ then $W_{0}$ approaches the ground--state energy $E_{0}$
as $N$ increases.

We have considered the same set of values of $\omega $, $\omega _{0}$ and $g$
used by Fessatidis et al.\cite{FMB02} in their application of the CMX to the
Rabi Hamiltonian. For each set of values we have obtained CMX results for $%
m\leq 49$ that requires up to $99$ connected moments. As we have
already mentioned, we do not expect to confirm the results of
Fessatidis et al\cite{FMB02}, even qualitatively, because their
approach based on only five moments is affected by an error in
$\mu _{3}$.

For concreteness we restrict the discussion to the case $\omega =\omega
_{0}=1$ and three values of $g$. The conclusions apply to the other cases as
well. In order to compare the rate of convergence of the RRK and CMX we
calculate $l_{W}=\log |W_{0}^{(n+1)}-W_{0}^{(n)}|$ where $W_{0}^{(n)}$ is
the approximate ground--state eigenvalue of the Rabi Hamiltonian calculated
with $n$ moments by means of either method.

\begin{figure}[H]
\bigskip
\par
\begin{center}
\includegraphics[width=10cm]{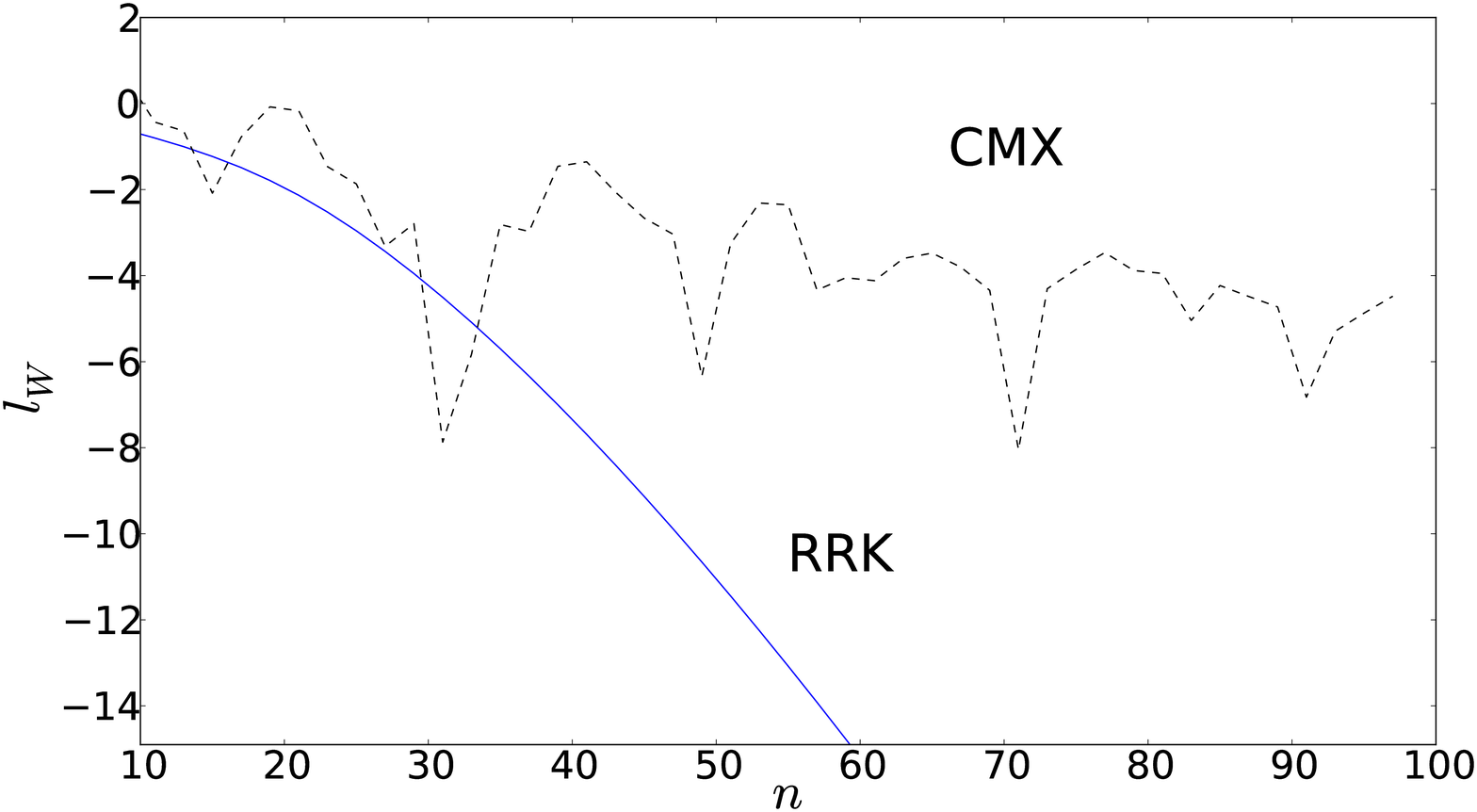}
\end{center}
\caption{Rate of convergence of the RRK (solid line) and CMX (dashed line)
for $\omega=1$, $\omega_0=1$, $g=1$}
\label{Fig:R111}
\end{figure}

\begin{figure}[H]
\bigskip
\par
\begin{center}
\includegraphics[width=10cm]{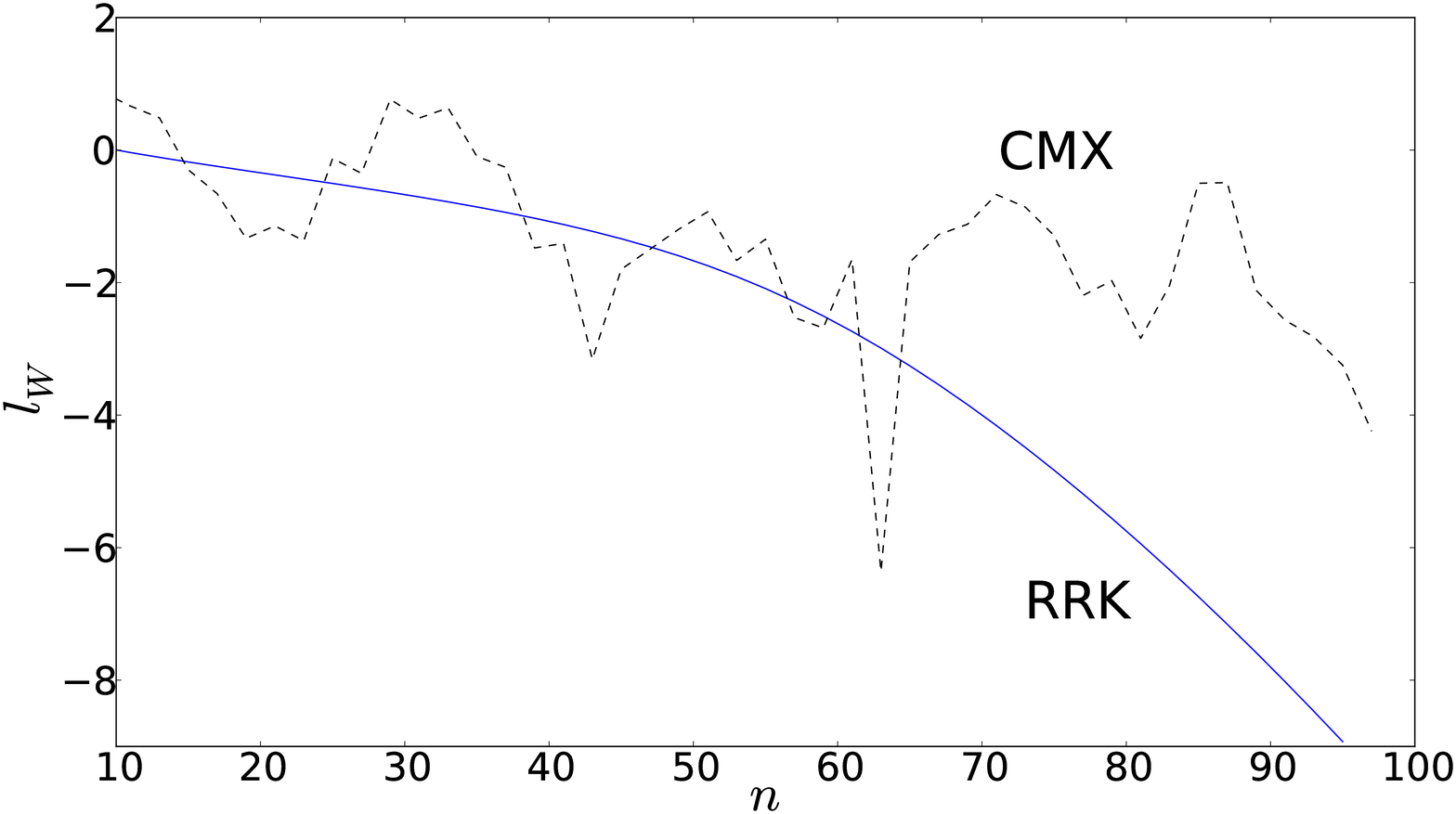}
\end{center}
\caption{Rate of convergence of the RRK (solid line) and CMX (dashed line)
for $\omega=1$, $\omega_0=1$, $g=2$}
\label{Fig:R112}
\end{figure}

\begin{figure}[H]
\bigskip
\par
\begin{center}
\includegraphics[width=10cm]{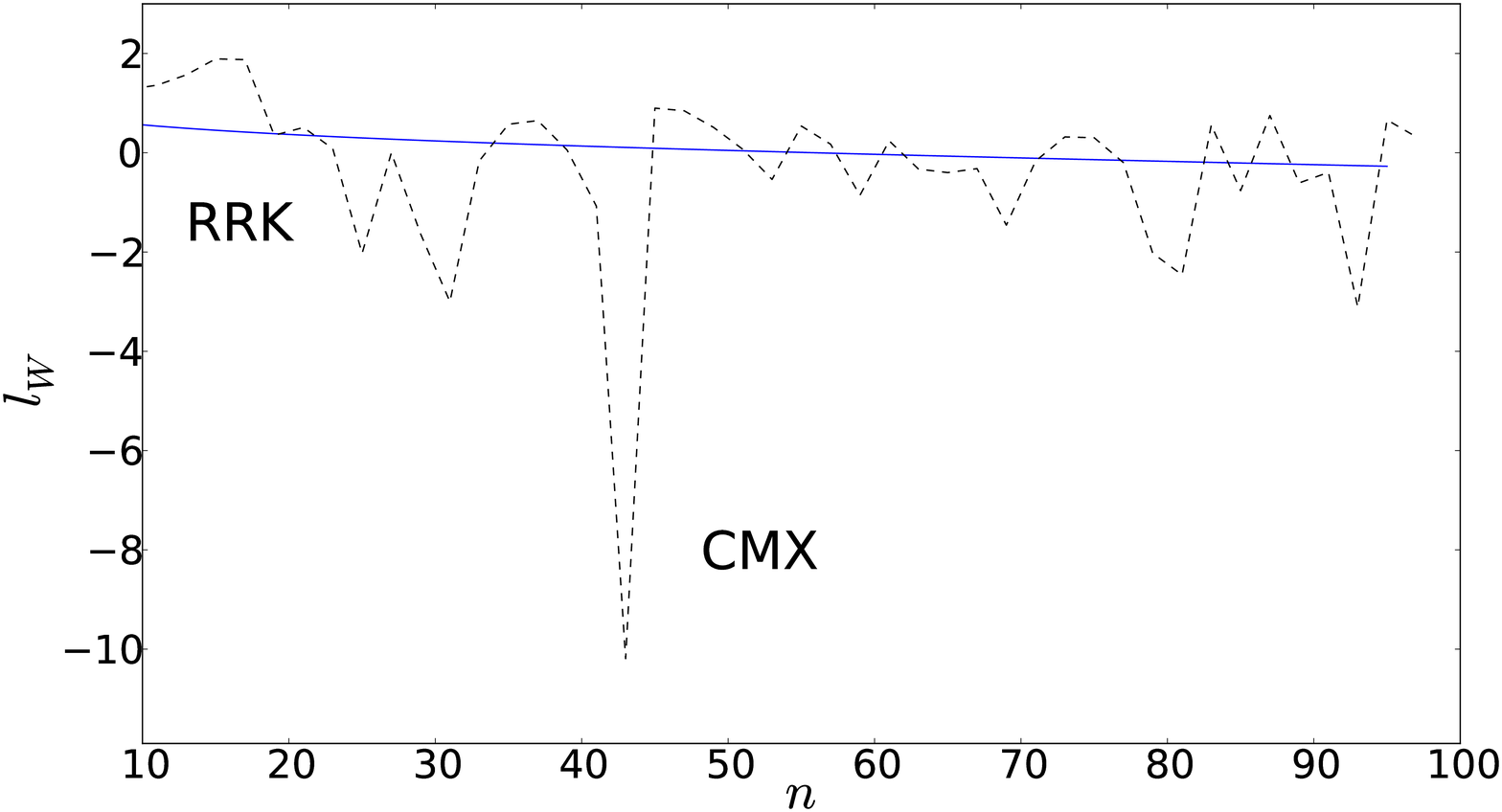}
\end{center}
\caption{Rate of convergence of the RRK (solid line) and CMX (dashed line)
for $\omega=1$, $\omega_0=1$, $g=5$}
\label{Fig:R115}
\end{figure}

Figures~\ref{Fig:R111} and \ref{Fig:R112} show that the RRK converges faster
and more smoothly than the CMX for $g=1$ and $g=2$, respectively. When $g=5$
the rate of convergence of the RRK is extremely slow. However, although the
CMX appears to give better results at some orders, the great oscillations of
$l_{W}$ render this method rather unreliable as shown in Fig.~\ref{Fig:R115}%
. Those figures look quite similar to plots of $\log |W_{0}^{(n)}-E_{0}|$
vs. $n$ and are therefore reasonable estimates of the rate of convergence of
the moments methods.

\section{Conclusions}

\label{sec:comments}

Some time ago the moments expansions appeared to be a promising
tool for the calculation of the ground--state energy of
quantum--mechanical problems of physical interest. However, some
judicious investigations suggested that the moments expansions are
unreliable and that in some cases they can even yield wrong
results\cite{K87}. In spite of this fact there is still some
unclear and inconclusive investigation on the convergence
properties of the moments expansions. With the purpose of adding
valuable information to that research we have tested the CMX on
the Rabi Hamiltonian that has already been treated by means of the
moments expansions\cite{FMB02}. We have calculated moments and
connected moments of much larger order than those considered
before. Our results clearly show that the CMX is unreliable
because the successive approaches to the ground--state energy
oscillate so strongly that one is never sure of the accuracy of
any particular calculation. Lee and Lo\cite{LL93} arrived at
similar conclusions for another model although by numerical
calculations of only fifth order.

We have contrasted the CMX with the RRK and showed that the latter converges
more smoothly from above towards the exact eigenvalues. It is reasonable to
compare these approaches that are based on the same kind of moments of the
Hamiltonian operator. Whereas the CMX provides an approximation to the
lowest eigenstate--eigenvalue of a given symmetry, the RRK yields all the
eigenstates--eigenvalues simultaneously because it is based on the
Rayleigh--Ritz variational method. The main advantage of the
connected--moments expansions, namely, size consistency\cite{HW84,K87}, is
not an issue in the case of simple problems like the one discussed here. In
other words: the RRK is preferable to the CMX and its variants in most (if
not all) the cases treated so far by means of the connected--moments
expansions.

In our opinion most standard approximation methods are more reliable than
the moments expansions.

\ack
P. Amore acknowledges support of Conacyt through the SNI
fellowship.

\end{document}